\documentclass[12pt,pre,aps,showpacs,preprint]{revtex4}

\usepackage{graphics}
\usepackage{graphicx}

\begin{document}

\title{Modeling and Reconstructing Complex Heterogeneous Materials
From Lower-Order Spatial Correlation Functions Encoding
Topological and Interface Statistics}

\author{Yang Jiao}

\email{yang.jiao.2@asu.edu}

\affiliation{Materials Science and Engineering, Department of
Physics, Arizona State University, Tempe, 85287, USA}

\date{\today}

\pacs{05.20.-y, 61.43.-j}

\begin{abstract}
The versatile physical properties of heterogeneous materials are
intimately related to their complex microstructures, which can be
statistically characterized and modelled using various spatial
correlation functions containing key structural features of the
material's phases. An important related problem is to inversely
reconstruct the material microstructure from limited morphological
information contained in the correlation functions. Here, we
present in details a generalized lattice-point (GLP) method based
on the lattice-gas model of heterogeneous materials that
efficiently computes a specific correlation function by updating
the corresponding function associated with a slightly different
microstructure. This allows one to incorporate the widest class of
lower-order correlation functions utilized to date, including
those encoding topological connectedness information and interface
statistics, into the Yeong-Torquato stochastic reconstruction
procedure, and thus enables one to obtain much more accurate
renditions of virtual material microstructure, to determine the
information content of various correlation functions and to select
the most sensitive microstructural descriptors for the material of
interest. The utility of our GLP method is illustrated by
modelling and reconstructing a wide spectrum of random
heterogeneous materials, including ``clustered'' RSA disks, a
metal-ceramic composite, a two-dimensional slice of a
Fontainebleau sandstone and a binary laser-speckle pattern, among
other examples.

\end{abstract}








\maketitle

\section{Introduction}


Heterogeneous materials (or random media) are those composed of
domains of different materials or phases or the same material in
different states. Such materials are ubiquitous in nature and in
man-made situations; examples include sandstones, concrete, animal
and plant tissue, gels and foams and distribution of galaxies
\cite{torquato, Sa03, Sa03II, Li00, Pe93, astrophysics, ecology,
Kh08, concrete}. Their versatile macroscopic (e.g., transport,
mechanical and electromagnetic) properties which are of great
interest in various engineering applications are intimately
related to the complex material microstructure \cite{torquato,
Sa03, Sa03II, To97}. Accordingly, a larger number of statistical
morphological descriptors have been devised to quantify the key
structure features of different material systems \cite{IMMI,
moment, poly1, poly2, poly3, jom}. On family of such descriptors
include the standard $n$-point correlation functions $S_n({\bf
x}_1,\cdots,{\bf x}_n)$ \cite{torquato}. In particular, $S_n$
gives the probability of simultaneously finding $n$ points with
positions ${\bf x}_1,\cdots,{\bf x}_n$, respectively, in one of
the phases of the media. Of particular interest are the
lower-order $S_n$ (such as $S_1$, $S_2$ and $S_3$), which have
been computed for various models of heterogeneous materials and,
as a result, excellent estimations of the effective properties of
these media have been obtained under certain situations
\cite{torquato}. Recently, lower order correlation functions have
been also been employed in computational material design schemes
\cite{adam_design, msea_design, northwestern_design}.

In the study of heterogeneous materials, an intriguing and
important inverse problem is the \textit{reconstruction} of these
media from a knowledge of limited microstructural information (a
set of lower-order correlation functions) \cite{reconstr1, Cu99,
Ji07, Ji08, utz, gaussian, gradient, FFT, multipoint, cross,
AIM_jiao, IMMI_jiao}. An effective reconstruction procedure
enables one to generate accurate digitized representations
(images) of the microstructure from lower-order correlation
functions obtainable in experiments or from theoretical
considerations, and subsequent analysis can be performed on the
images to obtain macroscopic properties of the material without
damaging the sample. Reconstruction of a three-dimensional medium
using information extracted from two-dimensional plane cuts
through the material is another application of great practical
value, especially in petroleum engineering, biology and medicine,
because in many cases only two-dimensional information such as a
micrograph is available. One can also determine how much
information is contained in the correlation functions by comparing
the original and reconstructed media. \textit{Construction} often
refers to generating realizations of heterogeneous materials from
a set of \textit{hypothetical} correlation functions, which
enables one to test the realizability of various types of
hypothetical functions -- an outstanding theoretical question
\cite{S2cond}. Recently, the (re)construction techniques have been
employed to identify and categorize heterogeneous materials based
on their correlation functions \cite{Ji07, Ji08} and to model a
wide spectrum of engineering materials, including sandstone
\cite{sandstone_sahimi, sandstone_sichun}, porous metal/cermaics
composite \cite{sofc}, alloys \cite{alloy_MMTA, alloy_JAP,
alloy_MaterChar, alloy_ActaMater}, and textile composites
\cite{texile_1, texile_2}. Very recently, correlation functions
have been incorporated to reconstruct complex multi-scale
polycrystal microstructure \cite{jiao_new01}, to fuse with limited
x-ray tomographic radiograph for material modeling
\cite{jiao_new02, jiao_new03, jiao_new04, jiao_new05, jiao_new06},
and to model microstructure evolution \cite{jiao_new07,
jiao_new08}.



A significant number of reconstruction studies focus on the
standard two-point correlation function $S_2({\bf r})$, which
gives the probability of finding two points separated by a
displacement vector ${\bf r}$ in the phase of interest. This
statistical descriptor can be obtained in small-angle X-ray
scattering experiments \cite{debye}. As pointed out in
Ref.~\cite{Ji09}, $S_2$ alone does not provide sufficient
information to uniquely determine the microstructures in general
\cite{deg1, deg2, deg3, deg4}. Although very recently, high
efficient methods have been developed to compute higher order
n-point correlation functions such as $S_3$ \cite{jiao_new09} and
to efficiently utilize partial higher order statistics in forms of
the n-point polytope functions \cite{jiao_new10}, the
computational costs associated with utilizing these higher order
statistics still limit their application in reconstructing large
3D realizations of complex material systems.

As noted in Ref. \cite{torquato}, in general one can not perfect
reconstruct the target microstructure using limited statistical
morphological information, i.e., such a reconstruction is
generally non-unique. Thus, the objective here is not the same as
that of data decompression algorithms which efficiently restore
complete information, but rather to generate realizations of
random microstructure with the key morphological features depicted
by the correlation functions. Instead of the aforementioned
natural and obvious extension to higher-order versions of $S_2$,
one could look at additional lower-order correlation functions
other than the standard $S_2$ for a better signature of the
microstructure.


In Ref.~\cite{Ji09}, we introduced a novel reconstruction
procedure called the generalized lattice-point (GLP) method, which
enables one to incorporate the widest class of lower-order
correlation functions examined to date. The GLP method allows one
to generate the accurate renditions of the media of interest using
various combinations of the correlation functions and to determine
the most sensitive statistical descriptors for the materials of
interest. Moreover we showed through several illustrative examples
in \cite{Ji09} that the two-point cluster function $C_2({\bf r})$,
which gives the probability of finding two points separated by a
displacement vector ${\bf r}$ in the \textit{same cluster}
\cite{cluster} of the phase of interest, is a superior statistical
descriptor to a variety of ``two-point'' quantities besides $S_2$,
including surface correlation functions $F_{ss}$ and $F_{sv}$, the
pore-size function $F$, lineal-path function $L$ and the
chord-length density function $p$ \cite{torquato} (all of which
are defined precisely in Sec.~2). However, the details of the GLP
method was not provided in Ref. \cite{Ji09}.


In this paper, we present the algorithmic details of the
generalized lattice-point method. In particular, the discretized
heterogeneous material is considered as a lattice-gas system
\cite{Ji08}, in which pixels with different local states are
``molecules'' of different ``gas'' species, or a {\it point}
process on a {\it lattice}. The correlation functions of interest
can be obtained by binning the separation distances between the
selected pairs of molecules from particular species. For
simplicity we only provide the formalism for binary random media
here. The generalization of the methodology to multi-phase
microstructures is straightforward. The GLP method is combined
with the Yeong-Torquato stochastic reconstruction technique
\cite{reconstr1} to evolve a trial microstructure to match the
specific target correlation functions as accurately as possible.
The GLP method is necessary to efficiently update the correlation
functions of the system during the reconstruction process to make
it computationally feasible to incorporate those functions into
the reconstruction: direct re-sampling is too computationally
expensive to implement in practice.

To demonstrate its utility, we apply the GLP method to reconstruct
a wide spectrum of random systems from a wide range of correlation
function, including ``clustered'' RSA disks, a metal-ceramic
composite, a two-dimensional slice of a Fontainebleau sandstone
and a binary laser-speckle pattern, among other examples. To
quantitatively ascertain the accuracy of a reconstruction,
correlation functions other than the targeted ones are measured
and compared to those of the original medium and the lineal-path
function $L(r)$ is used here. Except for the laser-speckle
pattern, which processes a multi-scale structure with percolating
phases, reconstructions incorporating $C_2$ always produce the
most accurate renditions of the target microstructures. This is
consistent with our conclusion in Ref.~\cite{Ji09} that
incorporation of $C_2$ significantly reduces the number of
compatible microstructures, even superior to certain higher order
n-point correlation functions. Statistical descriptors that could
be used to characterize multi-scale structures are also suggested.

The rest of the paper is organized as follows: In Sec.~2, we
define and discuss various correlation functions used in the
reconstructions. In Secs.~3 and 4, we provide the details of the
GLP method and how to incorporate it into the general stochastic
reconstruction procedure. In Sec.~5, we apply the methodology to
reconstruct a variety of random media. In Sec.~6, we make
concluding remarks.

\section{Definition of Correlation Functions}

Consider a $d$-dimensional two-phase (binary) microstructure in
which phase $i$ has volume fraction $\phi_i$ ($i=1,~2$) and is
characterized by the indicator function ${\cal I}^{(i)}({\bf x})$
defined as
\begin{equation}
\label{eq101} {\cal I}^{(i)}({\bf x}) = \left\{ {\begin{array}{*{20}c}
{1, \quad\quad {\bf x} \in V_i,}\\
{0, \quad\quad {\bf x} \in \bar{V_i},}
\end{array} }\right.
\end{equation}
where $V_i$ is the region occupied by phase $i$ and $\bar{V_i}$ is
the region occupied by the other phase. The two-point correlation
function is defined as
\begin{equation}
S^{(i)}_2({\bf x}_1,{\bf x}_2) = \left\langle{{\cal I}^{(i)}({\bf x}_1){\cal I}^{(i)}({\bf x}_2)}\right\rangle,
\end{equation}
where $\left\langle{\cdots}\right\rangle$ denotes ensemble
average. This correlation function is the probability of finding
two points ${\bf x}_1$ and ${\bf x}_2$  both in phase $i$.
Henceforth, we will drop the superscript ``$i$'' and only consider
the correlation functions for the phase of interest. For
\textit{statistically homogeneous} and \textit{isotropic}
microstructures, which is the focus of the rest of the paper,
two-point correlation functions will only depend on the distance
$r \equiv |{\bf x}_1-{\bf x}_2|$ between the points and hence
$S_2({\bf x}_1,{\bf x}_2)=S_2(r)$. In the absence of long-range
order, which is the most common occurrence, $S_2$ rapidly decays
to $\phi^2$, the probability of finding two points
\textit{independently} in the phase of interest.

The surface-void and surface-surface correlation functions are
respectively defined as
\begin{equation}
F_{sv}(r) =
\left\langle{{\cal M}({\bf x}_1){\cal I}({\bf x}_2)}\right\rangle, \qquad F_{ss}(r) = \left\langle{{\cal M}({\bf x}_1){\cal M}({\bf
x}_2)}\right\rangle,
\end{equation}
where ${\cal M}({\bf x})=|\nabla {\cal I}({\bf x})|$ is the
two-phase interface indicator function. By associating a finite
thickness with the interface, $F_{sv}$ and $F_{ss}$ can be
interpreted, respectively, as the the probability of finding ${\bf
x}_1$ in the ``dilated'' interface region and ${\bf x}_2$ in the
void phase and the probability of finding both ${\bf x}_1$ and
${\bf x}_2$ in the ``dilated'' interface region but in the limit
that the thickness tends to zero \cite{torquato}.

\begin{figure}
\includegraphics[width=0.45\textwidth,keepaspectratio]{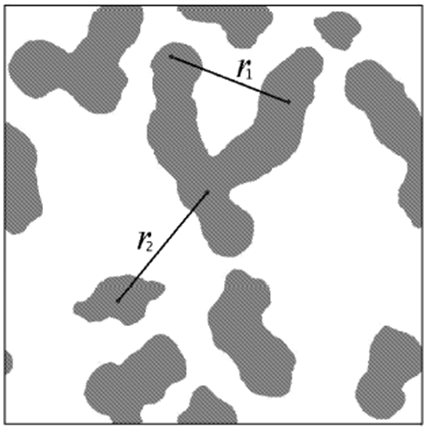}
\caption{Probability interpretation of $C_2$ and $S_2$. The point
pair separated by $r_1$ contributes to both $C_2$ and $S_2$; the
point pair separated by $r_2$ only contributes to $S_2$.}
\label{fig1}
\end{figure}

The lineal-path function $L(r)$ is the probability that an entire
line of length $r$ lies in the phase of interest, and thus
contains a coarse level of {\it connectedness} information, albeit
only along a {\it lineal path} \cite{torquato,Lu92}. The
chord-length density function $p(r)$ is the probability density
function associated with finding a ``chord" \cite{chord} of length
$r$ in the phase of interest and is directly proportional to the
second derivative of $L(r)$ \cite{To93}. The pore-size function
$F(\delta)$ is related to the probability that a sphere of radius
$r$ can lie entirely in the phase of interest \cite{torquato}, and
therefore is the ``spherical" version of the lineal measure $L$.

The two-point cluster function embodies a greater level of
connectedness information than either $L$ or $F$. In particular,
$C_2(r)$ is defined to be the probability of finding two points
separated by a distance $r$ in the same {\it cluster} of the phase
of interest \cite{To88,cluster}, as schematically shown in
Fig.~\ref{fig1}. When the phase is not percolating, $C_2$ is short
ranged and decays to zero rapidly. As the size of the clusters in
the systems increases, $C_2$ becomes a progressive longer-ranged
function such that its volume integral diverges at the percolation
threshold \cite{torquato}. Thus $C_2$ is extremely sensitive to
topologically connectedness information and it takes into account
all possible connecting paths, not only the ``lineal'' and
``spherical'' ones.

\section{Sampling the Correlation Functions: the Generalized Lattice-Point
Method}

The aforementioned probabilistic interpretations of the
correlation functions enable us here to develop a general sampling
method for reconstruction of statistically homogeneous and
isotropic digitized microstructures using the ``lattice-gas"
formalism described in detail in Ref.~\cite{Ji08}. In the
``lattice-gas'' formalism, the pixels have two local states, i.e.,
they are either hard ``lattice-gas molecules'' or unoccupied
lattice sites. Here we generalize the formalism to include
multiple local states and ``gas'' species: different pixel values
correspond to distinct local states and pixels with the same value
are considered to be ``molecules'' of the same ``gas'' species.
The correlation functions of interest can be obtained by binning
the separation distances between the selected pairs of molecules
from particular species.

\subsection{Standard Two-Point Correlation Function}

We denote the number of lattice-site separation distances of
length $r$ by $N_S(r)$ and the number of molecule-pair separation
distances of length $r$ by $N_P(r)$. Thus, the fraction of pair
distances with both ends occupied by the phase of interest, i.e.,
the two-point correlation function, is given by $S_2(r) =
N_P(r)/N_S(r)$.

\subsection{Two-Point Cluster Function}

To obtain $C_2$, one needs to partition the ``molecules'' into
different subsets $\Gamma_i$ (``species'') such that any two
molecules of the same species are connected by a path composed of
the same kind of molecules, i.e., molecules that form a cluster,
which is identified using the ``burning'' algorithm
\cite{burning}. The number of pair distances of length $r$ between
the ``molecules'' within the same subset $\Gamma_i$ is denoted by
$N^i_P(r)$. The two-point cluster function is then given by
$C_2(r) = \sum_i N^i_P(r)/N_S(r)$.

\subsection{Surface Correlation Functions}

The calculation of $F_{ss}$ and $F_{sv}$ requires partitioning the
``molecules'' into two subsets: the surface set $\kappa_S$
containing only the ``molecules'' on the surfaces of the clusters
and the volume set $\kappa_V$ containing the rest. In a digitized
medium, the interface necessarily has a small but finite thickness
determined by the pixel size. Thus, the surface-surface and
surface-void correlation functions can be regarded as
probabilities that are given by $F_{ss} = N^{ss}(r)/N_S(r)$ and
$F_{sv} = N^{sv}(r)/N_S(r)$, respectively; where $N^{ss}(r)$ gives
the number of distances between two surface molecules with length
$r$ and $N^{sv}$ is the counterpart for pairs with one molecule on
the surface and the other inside the cluster.

\subsection{Pore-size Function and Lineal-Path Function}

The pore-size function $F(\delta)$ can be obtained by integrating
the pore-size probability density function $f(\delta)$, which
provides the distribution of the maximal distance from a randomly
selected point in the ``pore'' phase to the two-phase interface.
We note here that ``pore'' phase is just used to refer a generic
phase which is necessarily to be void. To compute the probability
density function $f$ (and thus the pore-size function $F$), the
pixels are again partitioned into the surface set $\kappa_S$ and
the volume set $\kappa_V$. Then the maximal distance from each
pixel in $\kappa_V$ to the associated boundary pixel in $\kappa_S$
is computed and a histogram is generated to obtain $f$. A
numerical integration of $f$ will lead to the pore-size function
$F$.

The lineal path function $L$ (and hence the chord-length density
$p$) is computed by only sampling along orthogonal directions
consistent with the underlying lattice of the digitized
microstructure. This method has been well documented in literature
\cite{reconstr1, sandstone}, and hence will not be elaborated
here.

\section{Stochastic Microstructure Reconstruction}

\subsection{The Yeong-Torquato Procedure}

The stochastic optimization reconstruction procedure introduced by
Yeong and Torquato \cite{reconstr1} is ideally suited here because
it can incorporate any types of statistical descriptors. The
Yeong-Torquato procedure has become a popular reconstruction
technique \cite{Ma99, ApplyA, ApplyB, ApplyC, ApplyD} because it
is both robust and simple to implement. Consider a given set of
correlation functions $f^{\alpha}_n({\bf r}_1,{\bf r}_2,...,{\bf
r}_n)$ of the phase of interest that provides partial information
on the microstructure of the medium. The superscript $\alpha$
denotes the type of correlation functions, and the subscript $n$
denotes the order. The information contained in $f^{\alpha}_n$
could be obtained either from experiments or it could represent a
hypothetical medium based on simple models. In both cases we would
like to recover or generate the underlying microstructure. In the
former case, the formulated inverse problem is frequently referred
to as a ``reconstruction'' procedure, and in the latter case as a
``construction''. The (re)construction problem can be formulated
as an optimization problem in which the discrepancies between the
statistical properties of the generated structure and the imposed
ones are minimized. This is achieved by introducing the ``energy''
function $E$, defined as a sum of squared differences between
target correlation functions which we denote by
$\widehat{f}^{\alpha}_n$, and those calculated from generated
structures, i.e.,

\begin{equation}
\label{eq208}
E = \sum\limits_{{\bf r}_1,{\bf r}_2,...,{\bf r}_n}\sum\limits_{\alpha}\left[{f^{\alpha}_n({\bf r}_1,{\bf r}_2,...,{\bf r}_n)-\widehat{f}^{\alpha}_n({\bf r}_1,{\bf r}_2,...,{\bf r}_n)}\right]^2.
\end{equation}

\noindent For every generated structure (configuration), there is
a set of corresponding $f^\alpha_n$. If we consider every
structure (configuration) as a ``state'' of the system, $E$ can be
considered as a function of the states. The optimization technique
suitable for the problem at hand is the method of simulated
annealing, the concept of which is based on a well-known physical
fact: if a system is heated to a high temperature $T$ and then
sufficiently slowly cooled down to absolute zero, the system
equilibrates to its ground state. At a given temperature $T$, the
probability of being in a state with energy $E$ is given by the
Boltzmann distribution $P(E) \sim \exp(-E/T)$. At each annealing
step $k$, the system is allowed to evolve long enough to
equilibrate at $T(k)$. The temperature is then lowered according
to a prescribed annealing schedule $T(k)$ until the energy of the
system approaches its ground state value within an acceptable
tolerance. It is important to keep the annealing rate slow enough
in order to avoid trapping in some metastable state.

For our problem, the discrete configuration space includes the
states of all possible pixel allocations. Starting from a given
state (current configuration) with energy $E_{old}$, a new state
(trial configuration) can be obtained by randomly moving an
arbitrarily selected pixel (``gas molecule'') of the phase of
interest. This simple evolving procedure preserves the volume
fraction of all involved phases and guarantees ergodicity in the
sense that each state is accessible from any other state by a
finite number of steps. However, in the later stage of the
procedure, biased and more sophisticated pixel (``molecule'')
selection rules, i.e., surface optimization, could be used to
improve the efficiency \cite{Ji08}. Then the correlation functions
of the trial configuration are computed to obtain the energy
$E_{new}$ of this new state. Whether the trial configuration will
be accepted or rejected is based on the Metropolis criterion: the
acceptance probability $P$ is given by

\begin{equation}
\label{eq209}
P(E_{old}\rightarrow E_{new}) = \left\{{\begin{array}{*{20}c}1, \quad\quad \Delta E < 0, \\\\ \exp(-\Delta E/T),\quad \Delta E\ge 0,\end{array}}\right.\\
\end{equation}

\noindent where $\Delta E = E_{new}-E_{old}$. The temperature $T$
is initially chosen so that the initial acceptance probability of
trial configurations with $\Delta E\ge 0$ averages approximately
$0.5$. An inverse logarithmic annealing schedule which decreases
the temperature according to $T(k) \sim 1/\ln(k)$ would in
principle evolve the system to its ground state. However, such a
slow annealing schedule is difficult to achieve in practice. Thus,
we will adopt the more popular and faster annealing schedule
$T(k)/T(0) = \lambda^k$, where constant $\lambda$ must be less
than but close to unity. This may yield suboptimal results, but,
for practical purposes, it will be sufficient. The convergence to
an optimum is no longer guaranteed, and the system is likely to
freeze in one of the local minima if the thermalization and
annealing rate are not adequately chosen.

\subsection{Cluster and Surface Events: Efficient Updating Correlation Functions}

The aforementioned stochastic reconstruction procedure requires
generating and sampling a large number of configurations. The
efficient and isotropic sampling method introduced in Sec.~3.A
also enables one to quickly re-compute the desired correlation
functions of the new configuration based on the old ones, and thus
make the incorporation of those functions feasible: direct
re-sampling is too computationally expensive to implement in
practice. In this section, we present the details for efficiently
updating the correlation functions.

A distance matrix $D_{ij}$ storing the separation distances of all
``molecule'' pairs is established when the system is initialized
and the ``molecules'' are partitioned into different ``species''
depending on their positions. The quantities $N_P, N^i_P, N^{ss}$
and $N^{sv}$ can be obtained by binning the distances of
corresponding pairs. For $N_P$ one simply bins the distances of
all ``molecule'' pairs. For example, for $N^i_P$ one needs to
consider only the distances between two ``molecules'' belonging to
the same cluster; for $N^{sv}$ and $N^{ss}$ the distances between
``molecule'' pairs in which one ``molecule'' belongs to the
surface set and the other belongs to the volume set, and the
distances between ``molecules'' on the surface, respectively, are
considered. The motion of the randomly chosen ``molecule'' could
result in changes of the distances between it and all the other
``molecules'' as well as two kinds of ``species'' events. The
first kind involves breaking and combining clusters: if the chosen
``molecule'' happens to be the ``bridge'' between two
sub-clusters, removing the ``bridge'' would make the original
single cluster break into two pieces, i.e., a new ``species'' is
generated; similarly, the reverse of the above process would lead
to combination of clusters and annihilation of ``species''. The
other kind of ``species'' event is the transition of ``molecules''
between surface and volume ``species'': if a surface ``molecule''
is removed, certain volume ``molecules'' would now constitute the
new surface and vice visa; the chosen ``molecule'' could also
undergo such transitions depending on its original and new
positions, e.g., an volume ``molecule'' originally inside a
cluster could be moved outside and becomes a surface ``molecule''.

The old contributions of the number of distance pairs to $N_P,
N^i_P, N^{ss}$ and $N^{sv}$ from the ``molecules'' undergone the
``species'' transitions due to the moved ``molecule'' are computed
and subtracted accordingly; the new contributions can be obtained
by binning the distances of ``molecule'' pairs belonging to
particular new ``species'' and are added to the corresponding
quantities. This method only requires operations on a finite small
number of ``molecules'', including retrieving and binning their
separation distances and updating the ``species'' sets. The
distance matrix $D_{ij}$ would speed up the former process,
however for very large systems (including millions of
``molecules'') storing $D_{ij}$ requires very large memory space
on a computer. An alternative is to recompute the distances
involving the ``molecules'' undergone the ``species'' transitions
for every trial configuration instead of explicitly storing all
distances. This may slightly slow down the process but make it
easy to handle very large systems. Correlation functions of the
new configuration could be obtained by dividing updated $N_P,
N^i_P, N^{ss}$ and $N^{sv}$ by $N_S$, as discussed in Sec.~3.A.
The complexity of the algorithm is linear in the total number of
`` molecules'' within the system.

\section{Applications}


The generalized lattice-point method has been employed to
reconstruct a wide spectrum of statistically homogeneous and
isotropic random media, including both theoretical model
microstructures \cite{torquato} and digitized representations of
real materials. Each  of the two-point functions considered
depends only on scalar distance. In the following, we will present
in details several reconstruction examples to illustrate the
utility of the GLP method, which include ``clustered'' RSA disks
(explained below), a metal-ceramic composite \cite{ceramic}, a
two-dimensional slice of a Fontainbleau sandstone \cite{sandstone}
and a laser-speckle pattern \cite{Ji08}.

Reconstructions from various combinations of the standard
two-point correlation function $S_2(r)$, the surface correlation
functions $F_{ss}(r)$ and $F_{sv}(r)$ and the two-point cluster
function $C_2(r)$ are shown here. (Since the pore-size function
$F(\delta)$ only contain partial connectedness information,
reconstructions based on $F$ is expected to be inferior to those
incorporated $C_2$.) For any target configuration, one must choose
different cooling schedules for different choices of the set of
target correlation functions in order to achieve the same final
energy $E$ (or error), which is of order $10^{-8}$-$10^{-11}$,
depending on the example. The accuracy of a reconstruction can be
ascertained quantitatively by measuring correlation functions
other than the targeted ones and comparing the other correlation
functions to those of the original medium. Here we measure the
lineal-path function $L(r)$ of each reconstruction and compare
with that sampled from the target media. We will see the results
clearly indicate that when the phase of interest is not
percolating, $C_2(r)$ not only contains appreciable more
microstructural information than $S_2$, but more than a variety of
other ``two-point'' quantities.

\subsection{``Clustered'' RSA Disks}

\begin{figure}[bthp]
\includegraphics[width=0.85\textwidth,keepaspectratio]{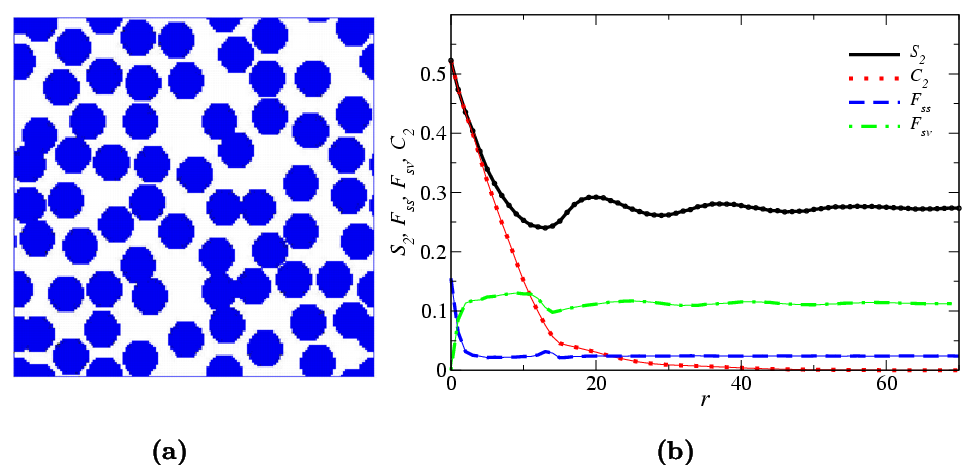}
\caption{(color online). (a) Target configuration: a ``clustered''
RSA disk system as described in the text. The volume fraction of
the particle phase (shown in blue) is $\phi_1 = 0.523$ and the
volume fraction of the void phase (shown in white) is $\phi_2 =
0.477$. (b) The correlation functions of target and reconstructed
systems: various tick curves correspond to the target functions;
the thin solid curves correspond to the reconstructed functions. }
\label{fig_rsa_I}
\end{figure}

Fig.~\ref{fig_rsa_I} shows a two-dimensional realization of a
packing circular disks and the target correlation functions of the
particle phase (shown in blue). To obtain the target system in
Fig.~\ref{fig_rsa_I}(a), a random sequential addition (RSA)
distribution of hard disks is generated \cite{torquato} and then
several randomly chosen disks are brought into proximity with one
another to form complex ``clusters'', and thus are called
``clustered'' RSA disks.

\begin{figure}[bthp]
\includegraphics[width=0.95\textwidth,keepaspectratio]{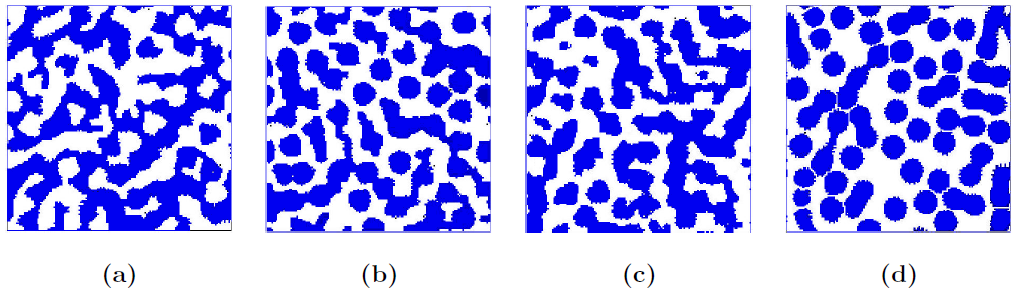}
\caption{(color online). Reconstructed ``clustered'' RSA disk systems:
(a) $S_2$-alone reconstruction. (b) $S_2$-$F_{ss}$ reconstruction. (c) $S_2$-$F_{sv}$ reconstruction.
(d) $S_2$-$C_2$ reconstruction. }
\label{fig_rsa_F}
\end{figure}

\begin{figure}
\includegraphics[width=0.55\textwidth,keepaspectratio]{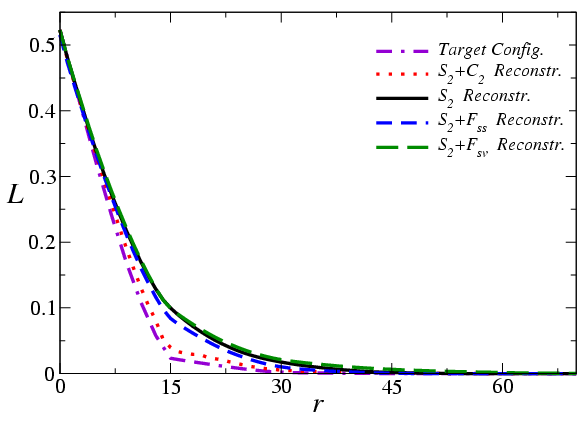}
\caption{(color online). The lineal-path function $L(r)$ of the
target and reconstructed systems.} \label{fig_rsa_L}
\end{figure}

Fig.~\ref{fig_rsa_F} shows that the reconstruction using $S_2$
alone overestimates clustering in the system and incorrectly
yields a percolating ``particle'' phase. Thus, although $S_2$ of
the ``particle'' phase of the reconstructed realization  matches
that of the target one with very small error ($E \sim 10^{-9}$),
such information is insufficient to get a good reconstruction.
Incorporating both $S_2$ and surface correlation functions
$F_{ss}$ or $F_{sv}$ leads to
 better renditions of the target system but
the reconstructions still overestimate the degree of clustering.
On the other hand, incorporating
$C_2$ of the particle phase yields the best reconstruction. Although the latter
is still not perfect, it is clear that it has the essential features
of the actual dispersion, such as the presence of individual
circular disks as well as small clusters.
The lineal-path functions of the target and reconstructed systems
are shown in Fig.~\ref{fig_rsa_L}. Again, $L$ of the $S_2$-$C_2$
hybrid reconstruction matches the target lineal-path function (unconstrained in the reconstruction)
much better than the others.

\subsection{Metal-Ceramic Composite}

\begin{figure}[bthp]
\includegraphics[width=0.85\textwidth,keepaspectratio]{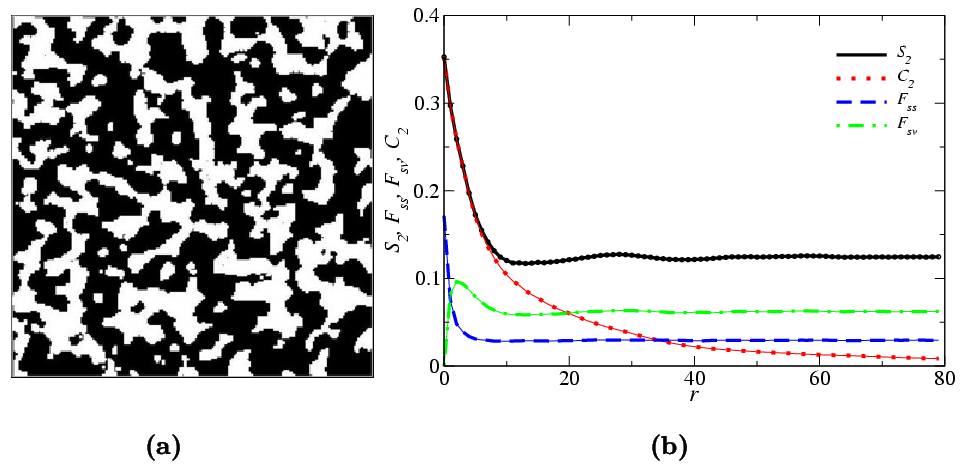}
\caption{(color online). (a) Target configuration: a digitized
image of a boron-carbide/aluminum composite. The black phase is
boron carbide with $\phi_1 = 0.647$ and the white phase is
aluminum with $\phi_2 = 0.353$. (b) The correlation functions of
target and reconstructed configurations: various tick curves
correspond to the target functions; the thin solid curves
correspond to the reconstructed functions. } \label{fig_ceramic_I}
\end{figure}

\begin{figure}[bthp]
\includegraphics[width=0.95\textwidth,keepaspectratio]{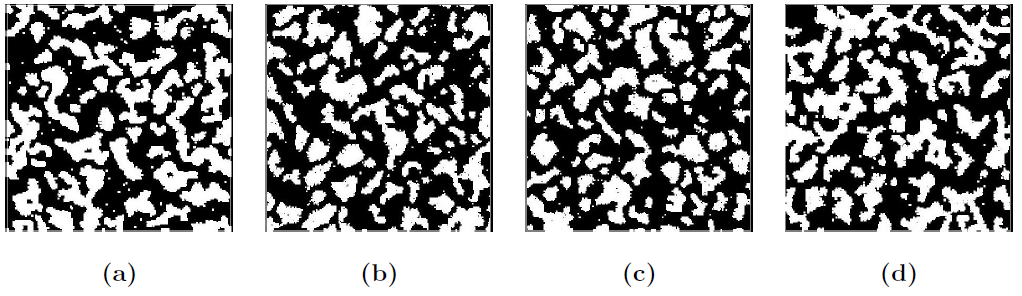}
\caption{Reconstructed configurations of the
boron-carbide/aluminum composite: (a) $S_2$-alone reconstruction.
(b) $S_2$-$F_{ss}$ reconstruction. (c) $S_2$-$F_{sv}$
reconstruction. (d) $S_2$-$C_2$ reconstruction. }
\label{fig_ceramic_F}
\end{figure}

\begin{figure}
\includegraphics[width=0.55\textwidth,keepaspectratio]{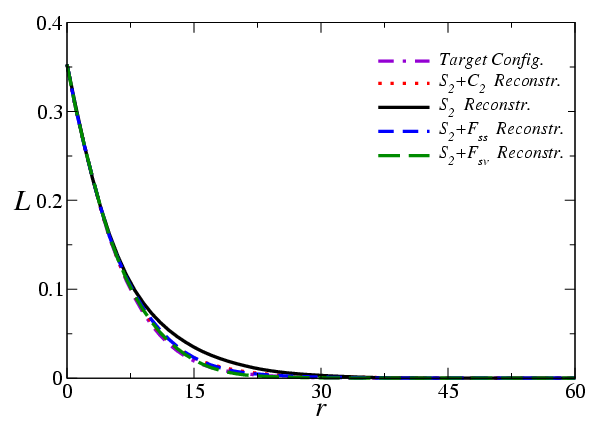}
\caption{(color online). The lineal-path function $L(r)$ of the
target and reconstructed configurations.} \label{fig_ceramic_L}
\end{figure}

A two-dimensional digitized image of a boron-carbide/aluminum
(B$_4$C/Al) interpenetrating composite and the target correlation
functions of the aluminum phase (shown in white) are shown in
Fig.~\ref{fig_ceramic_I} \cite{Ji08, ceramic}.
Fig.~\ref{fig_ceramic_F} shows the reconstructions, which all well
match the target image visually. However, the reconstruction
incorporating $C_2$ has the a minimum sum of absolute
discrepancies between the measured and target lineal-path
functions (shown in Fig.~\ref{fig_ceramic_L}), indicating it is
superior to other statistical descriptors utilized. The
$S_2$-alone reconstruction again overestimates the clustering in
the system, leading to larger values of $L$ for intermediate range
of $r$.

\subsection{Fontainebleau Sandstone}

\begin{figure}[bthp]
\includegraphics[width=0.85\textwidth,keepaspectratio]{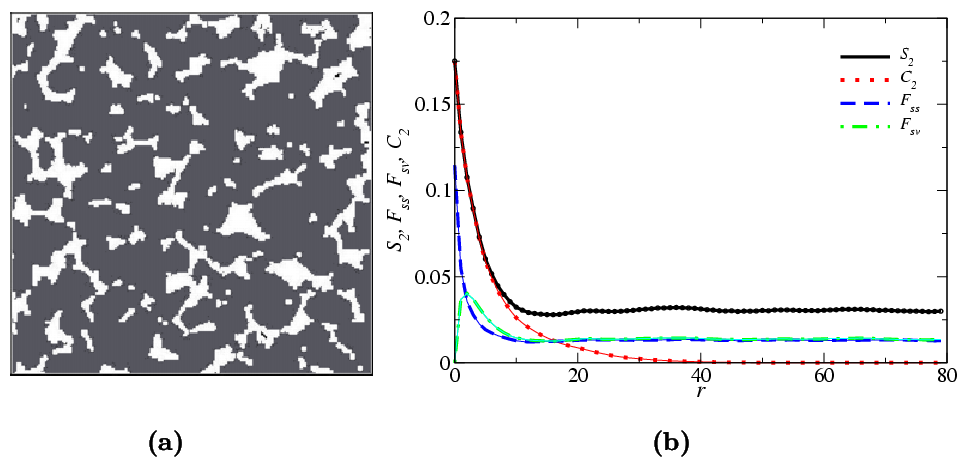}
\caption{(color online). (a) Target configuration: a
microstructural image of a slice of a Fontainebleau sandstone. The
volume fraction of the rock phase (shown in gray) if $\phi_1 =
0.825$ and the volume fraction of the porous phase (shown in
white) is $\phi_2 = 0.175$. (b) The correlation functions of
target and reconstructed configurations: various tick curves
correspond to the target functions; the thin solid curves
correspond to the reconstructed functions. }
\label{fig_sandstone_I}
\end{figure}

Fig.~\ref{fig_sandstone_I}(a) shows a tomographic image of a
two-dimensional (2D) slice of a Fontainebleau sandstone
\cite{sandstone, Ma00}. Sandstone is a porous material that has
important applications in geophysical science and petroleum
engineering. The volume fraction (porosity) and the topology of
the pore phase (shown in white) are crucial to the physical
properties of the sandstone, such as fluid permeability and
relaxation times. The correlation functions of the pore phase is
shown in Fig.~\ref{fig_sandstone_I}(b).

\begin{figure}[bthp]
\includegraphics[width=0.95\textwidth,keepaspectratio]{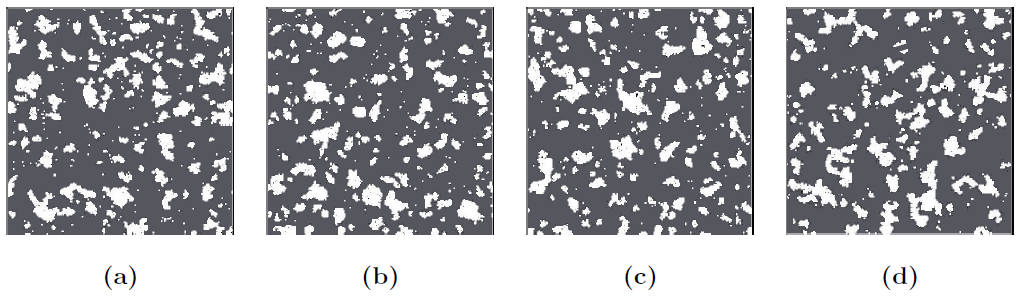}
\caption{Reconstructed configurations of the slice of the
Fontainebleau sandstone: (a) $S_2$-alone reconstruction. (b)
$S_2$-$F_{ss}$ reconstruction. (c) $S_2$-$F_{sv}$ reconstruction.
(d) $S_2$-$C_2$ reconstruction. } \label{fig_sandstone_F}
\end{figure}

\begin{figure}
\includegraphics[width=0.55\textwidth,keepaspectratio]{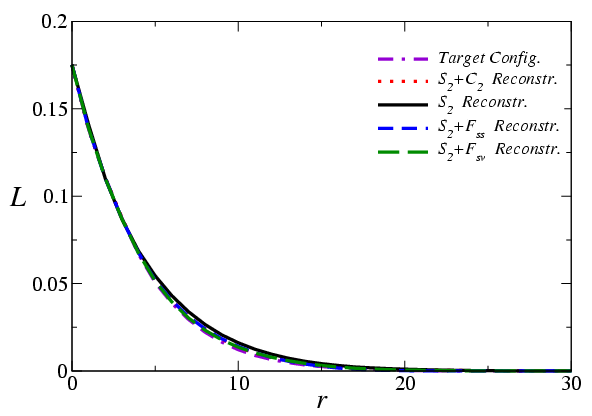}
\caption{(color online). The lineal-path function $L(r)$ of the
target and reconstructed configurations.} \label{fig_sandstone_L}
\end{figure}

The reconstructions are shown in Fig.~\ref{fig_sandstone_F}. In
two dimensions, the pore phase are topologically disconnected
regions enclosed by complicated concave contours. This feature is
best captured in the $S_2$-$C_2$ reconstruction. Although the
average size of the pore regions in other reconstructions also
match the target well, as can be seen from the comparison of the
lineal-path functions (see Fig.~\ref{fig_sandstone_L}), the their
shapes are not very satisfactory.

Note that even the discrepancies of the $S_2$-alone
reconstructions of the metal-ceramic composite and the
Fontainebleau sandstone are not that significant as in the first
two examples, which is consistent with our conclusion in
Ref.~\cite{Ji08} that $S_2$ is able to model the structures the
two materials to a high accuracy.

\subsection{Laser-Speckle Pattern}

\begin{figure}[bthp]
\includegraphics[width=0.85\textwidth,keepaspectratio]{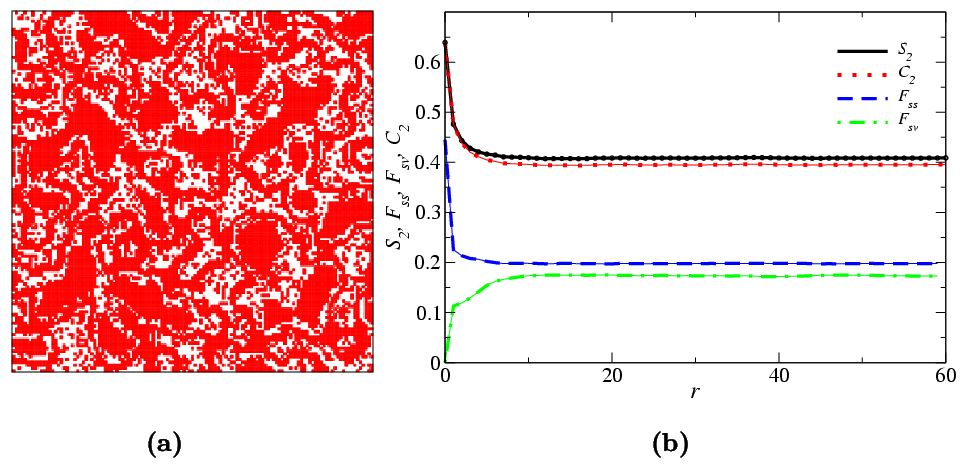}
\caption{(color online). (a) Target configuration: a digitized
image of a binary laser-speckle pattern. The volume fraction of
the red phase if $\phi_1 = 0.639$, and the volume fraction of the
white phase is $\phi_2 = 0.361$. (b) The correlation functions of
target and reconstructed configurations: various tick curves
correspond to the target functions; the thin solid curves
correspond to the reconstructed functions. } \label{fig_speckle_I}
\end{figure}

In all the aforementioned examples, the phases of interest are not
percolating so that $C_2$ could provide essential connectedness
information and lead to the best reconstructions. When percolating
clusters appear in the system, $C_2$ will become long ranged and
contain less additional structural information. In the limit that
all pixels belong to a single percolating cluster, $C_2$ and $S_2$
become identical.

In this section, we study a binary laser-speckle pattern
exhibiting a \textit{multi-scale structure}, in which both phases
are \textit{percolating} \cite{pixel_rule}. In Ref.~\cite{Ji08} we
pointed out that the structural information contained in $S_2$ is
averaged out over the several length scales of the pattern and the
$S_2$-alone reconstruction produced a typical Deybe-random
microstructure (containing clusters of ``random shapes and
sizes'') \cite{Ji07} associated with the exponentially decreasing
$S_2$, instead of generating a multi-scale pattern.

Fig.~\ref{fig_speckle_I}(b) shows the target correlation functions
of the red phase. It can be seen that $C_2$ is long-ranged and
almost identical with $S_2$ except for large $r$ values. This
indicates that the medium contains a single large cluster
composing most of the pixels; the discrepancies between $C_2$ and
$S_2$ are due to the ``background noise'', i.e., individual pixels
dispersed in the white phase. Also the information in $C_2$ is
averaged over the length scales. These could all make $C_2$ less
efficient than in the other examples.

\begin{figure}[bthp]
\includegraphics[width=0.95\textwidth,keepaspectratio]{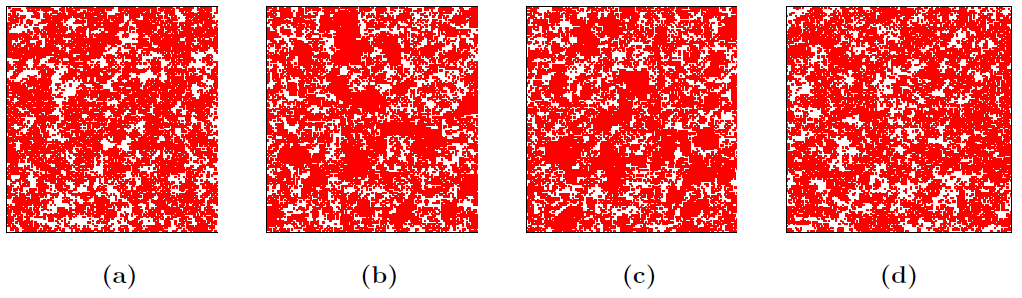}
\caption{(color online). Reconstructed configurations of the
laser-speckle pattern: (a) $S_2$-alone reconstruction. (b)
$S_2$-$F_{ss}$ reconstruction. (c) $S_2$-$F_{sv}$ reconstruction.
(d) $S_2$-$C_2$ reconstruction. } \label{fig_speckle_F}
\end{figure}

\begin{figure}
\includegraphics[width=0.55\textwidth,keepaspectratio]{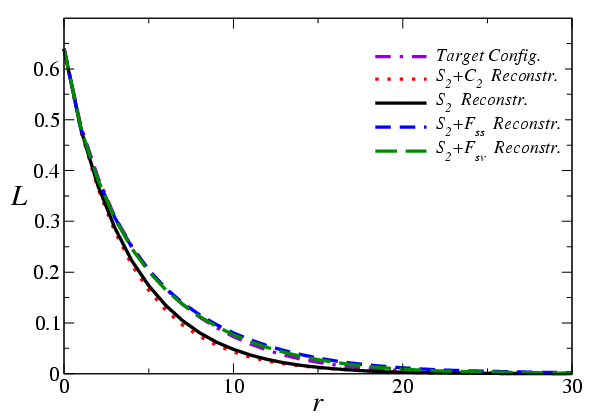}
\caption{(color online). The lineal-path function $L(r)$ of the
target and reconstructed configurations.} \label{fig_speckle_L}
\end{figure}

Indeed as we can see in Fig.~\ref{fig_speckle_F}, $S_2$-alone and
$S_2$-$C_2$ reconstructions produce similar microstructures of
Debye-random type. The reconstructions using surface correlation
functions are better in the sense that they contain compact
regions of similar sizes with those in the target medium, as can
be seen from the sampled lineal-path functions (see
Fig.~\ref{fig_speckle_L}). In other words, in the best
reconstruction of the laser-speckle pattern we obtained, only the
structures on the largest length scale could be resolved; the
filamentary structures connecting the compact clusters in the
target medium seem to mix up with the ``noise'' pixels in the
reconstructed media and are not observed.

We emphasize that the failure of $C_2$ in producing accurate
renditions of the target medium here is due to two reasons:
percolation of the phases and average of the length scales. The
former could be possibly resolved by defining different
pixel-connectivity rules. However, it is the latter makes the
medium extreme difficult to reconstruct. To efficiently
characterize multi-scale media, descriptors carrying non-trivial
information on different length scales are necessary.

\section{Conclusions}


In this paper, we provide the algorithmic details of the
generalized lattice-point method for heterogeneous material
reconstruction. The GLP method based on the lattice-gas model of
digitized random media enables one to incorporate the widest class
of lower-order correlation functions utilized to date into the
Yeong-Torquato stochastic reconstruction technique. Thus one could
use various combinations of those correlation functions to
generate renditions of the microstructures of interest with high
accuracy and determine the most sensitive statistical descriptors
for the material of interest. To illustrate its utility, we apply
the GLP procedure to reconstruct a wide spectrum of random media,
including ``clustered'' RSA disks, a metal-ceramic composite, a
two-dimensional slice of a Fontainebleau sandstone and a binary
laser-speckle pattern, among other examples. Except for the
laser-speckle pattern, which has a multi-scale microstructure with
percolating phases, reconstructions incorporating $C_2$ always
give the best renditions of the target medium. We note that a few
past investigations have recognized the need to use descriptors
containing connectedness information, such as the lineal-path
function $L$ \cite{reconstr1} and the pore-size distribution
function $F$ \cite{Ma99}. Those quantities only crudely reflect
this crucial topological information and the reconstructions using
$C_2$ are always found to be much more accurate than those
involving $L$ or $F$.


As we pointed out in \cite{Ji09}, $C_2$ is a superior
microstructural signature because it is extremely sensitive to
topologically connectedness information and becomes a
progressively longer-ranged function as clusters increase in size
in the system. By contrast, the quantities $S_2$, $L$, $F$,
$F_{ss}$ and $F_{sv}$ are insensitive to crossing the percolation
threshold. Thus, for statistically homogeneous and isotropic media
in which the phase of interest was below its percolation threshold
and contains essentially compact clusters of that phase,
incorporation of $C_2$ in reconstructions of such media could
provide renditions of the target structure with heretofore
unattained accuracy.

The unsuccessful reconstructions of the laser-speckle pattern are
mainly because that structural information on different length
scales are averaged out, and currently only the structures on the
largest scale can be resolved. To capture the salient features of
the multi-scale structures, statistical descriptors containing
information on different length scales are necessary. The
fluctuations of local properties, such as local volume fraction
\cite{localfluc} or surface area are possible candidates, which
give the fluctuations of local properties for \textit{different
sized} observation windows. Interestingly, our preliminary
calculations show that these quantities could indeed capture
structural features on different length scales.


Although we focused on the \textit{reconstruction} problems here,
the potential applications of the new method in the
\textit{constructions} should not be underestimated. Besides
$S_2$, the properties of other ``two-point'' function are less
well studied. The conditions (both necessary and sufficient) for a
hypothetical correlation function to be realizable, i.e., whether
it corresponds to a particular microstructure, have not been
systematically considered. Progress in these related topics would
open new doors for many fruitful applications. Our reconstruction
technique is ideally suitable to explore the nature of these
``two-point'' functions and would lead to a better understanding
of their properties. Moreover, given a particular correlation
function such as the two-point correlation function, one could
generate a family of microstructures with identical $S_2$ but
different $C_2$ or $F_{sv}$, etc., which could lead to development
of new schemes to model complex microstructures.

\begin{acknowledgments}
This work was supported by the Division of Materials Research at
the National Science Foundation under award No. DMR-1305119 (Dr.
Diana Farkas, program manager). Y. J. also thanks Arizona State
University for the generous start-up funds.
\end{acknowledgments}


\begin{thebibliography}{12}
\bibitem{cluster}
A cluster of phase $i$ is defined as the part of phase
 $i$ that can be reached from a point in phase $i$ without
passing through phase $j \neq i$.
\bibitem{chord}
Chords are the line segments between the intersections of an
infinitely long line with the two-phase interface.
\bibitem{pixel_rule}
The pixels are considered connected if they meet at the corners. Thus it
is possible to have bi-percolation even in two dimensions.
\end{thebibliography}
\end{document}